\documentstyle[12pt,aasms4]{article}
\lefthead{HATTORI ET AL.}
\righthead{DISCOVERY OF X-RAY EMISSION FROM CL2236-04}
\begin{document}
\title{Discovery of X-ray emission
from the distant lensing cluster of galaxies CL2236-04
at $z = 0.552$}
\author{{\sc Makoto Hattori}\altaffilmark{1,2,3},
{\sc Hideyuki Matuzawa}\altaffilmark{3,4},
{\sc Kohji Morikawa}\altaffilmark{1},
{\sc Jean-Paul Kneib}\altaffilmark{5},
{\sc Kazuyuki Yamashita}\altaffilmark{6},
{\sc Kazuya Watanabe}\altaffilmark{7},
{\sc Hans B\"ohringer}\altaffilmark{2},
{\sc and Takeshi G. Tsuru}\altaffilmark{8}}
\altaffiltext{1}{Astronomical Institute, T\^ohoku University, 
Aoba, Sendai 980-77, Japan;
hattori@astr.tohoku.ac.jp;morikawa@astr.tohoku.ac.jp}
\altaffiltext{2}{Max-Planck-Institut f\"ur extraterrestrische Physik,
Giessenbachstr., 85740 Garching, Germany;}
\altaffiltext{3}{The Institute of Physical and Chemical Research (Riken),
Hirosawa Wako, Saitama 351-01, Japan}
\altaffiltext{4}{Miyazaki Prefectural Nursing University, Gujibun Komogasako Otsu 2203, Miyazaki, 880, Japan;matuzawa@astro.miyazaki-u.ac.jp}
\altaffiltext{5}{Observatoire Midi-Pyr\'en\'ees, Laboratoire d'Astrophysique,
UMR 5574, 14 Avenus E. Belin, F-31400 Toulouse, France;kneib@obs-mip.fr}
\altaffiltext{6}{Information Processing Center, Chiba University, 
Chiba 263, Japan;yamasita@tanpopo.ipc.chiba-u.ac.jp}
\altaffiltext{7}{Department of Physics, Niigata University, 
Igarashi, Niigata 950-2102, Japan;kazuya@astro.sc.niigata-u.ac.jp}
\altaffiltext{8}{Department of Physics, Kyoto University,
Sakyo-ku, Kyoto, Japan;tsuru@cr.scphys.kyoto-u.ac.jp}
\begin{abstract}
X-ray emission from the distant lensing cluster CL2236-04 at $z$ = 0.552
was discovered by ASCA and ROSAT/HRI observations.
If the spherical symmetric mass distribution
model of the cluster is assumed,
the lensing estimate of the cluster mass is
a factor of two higher than that obtained from X-ray observations as reported for
many distant clusters.
However, the elliptical and clumpy lens model proposed by Kneib {\it et al.}(1993)  
is surprisingly consistent with 
the X-ray observations assuming that the X-ray emitting hot gas is
isothermal and in a hydrostatic equilibrium state.   
The existence of 
the cooling flow in the central region of  the cluster is 
indicated by the short central cooling time and the excess 
flux detected by ROSAT/HRI compared to the ASCA flux.  
However, it is shown that even if the AXJ2239-0429 has a cooling flow in the 
central region, the temperature measured by ASCA which is the mean emission-weighted cluster
temperature in this case, should not be cooler than and different from the virial temperature
of the cluster. 
Therefore, we conclude that the effect of the clumpiness  
and non-zero ellipticity in the mass distribution of the cluster 
are essential to explain the observed feature of the giant luminous
arc, and there is no discrepancy between strong lensing and
X-ray estimation of the mass of the cluster in this cluster.
\end{abstract}
\section{Introduction}
It is one of the prime objectives in observational cosmology
to determine mass distributions and total masses of clusters of galaxies
at various redshifts.
It  provides a clue
to constrain the unknown cosmological parameters:the species of dark matter,
the nature of cosmological mass density fluctuations, the average density in the universe
and the cosmological constant.\\
\indent
Recently, gravitational lensing has become a powerful tool
to measure the mass distribution in clusters.
Strong lensing events such as multiple images and/or giant luminous arcs
make it possible to model the central region of clusters.
However, many clusters show a mass discrepancy
with a  total mass deduced from strong lensing observation being
larger by a factor of 2--3
than that deduced from X-ray observations
(Loeb \& Mao 1994, Miralda-Escud\'e \& Babul 1995,  Kneib {\it et al.} 1995,
Schindler {\it et al.} 1995, \& 1997, Wu \& Fang 1997, Ohta, Mitsuda, Fukazawa 1997).
However, there are some clusters whose masses deduced in these two ways agree very well.
In such cases, the multi-phase of the intra-cluster gas (Allen, Fabian
\& Kneib 1996, Allen 1997)
or asymmetric mass distribution (Piere {\it et al.} 1996, 
B\"ohringer {\it et al.} 1997) has been taken into account
in the lens modeling.
This indicates that careful studies for a large number of
clusters are necessary before reaching
the conclusions that the lens modeling gives higher mass estimation
than that from X-rays.

CL2236-04 was discovered by Melnick {\it et al.} (1993) with a giant luminous arc.
The redshift of the arc was measured to be $z_{arc}$=1.116 (Melnick {\it et al.} 1993)
confirming the gravitational lens picture very satisfactorily.
The arc is rather straight and velocity structure across the arc is
measured (Melnick {\it et al.} 1993). 
These nature give strict constraints on the lensing mass estimation.
Further, CL2236-04 is one of the best and rare clusters at high redshift of $z>0.5$ for which
comparison of the mass distribution deduced by two individual methods is possible.
We have performed the first deep ASCA and ROSAT/HRI pointing observations toward CL2236-04.
In this paper, we report the discovery and the properties of X-ray emission from CL2236-04
and the mass distribution deduced from the X-ray observation is compared to the lensing mass.

\section{X-ray observations} 
\subsection{ASCA observation}
We performed an ASCA observation of CL2236-04 in June 1995 with an effective
exposure time of about 40ksec.
A new X-ray source was discovered in the direction of CL2236-04
with more than $10\sigma$ significance (hereafter referred  to as AXJ2239-0429).
The X-ray source count rate is $4\times 10^{-3}{\rm cts/sec}$ for each GIS
and $6\times 10^{-3}{\rm cts/sec}$ for each SIS.
The X-ray peak position of the source is
$(\alpha,\delta)_{2000}=(22^{\rm h}39^{\rm m}32^{\rm s},-4^{\rm d}30^{\rm m})$ (J2000).
This is consistent with the position of the cD galaxy,
$(\alpha,\delta)_{2000}=(22^{\rm h}39^{\rm m}32.767^{\rm s},-4^{\rm d}29^{\rm m}32.24^{\rm s})$
within the ASCA position determination  accuracy of
$(\Delta \alpha, \Delta \delta)=(\pm 4^{\rm s},\pm 1^{\rm m})$.

Figure 1 shows the X-ray spectra of AXJ2239-0429 obtained by
ASCA GIS2$+$3 and SIS0$+$1.
The SIS data were taken in 2CCD faint mode.
The background for GIS spectrum is taken from blank field observations during PV phase
of the ASCA mission where point sources contaminating the background spectrum were masked out.
The background for the SIS spectrum is taken from the same observation with masking the source region.
Assuming that the source is a cluster of galaxies at $z=0.552$,
the spectrum can be fit with a Mewe model modified by Galactic absorption.
We fixed the hydrogen column density of the galactic absorption to
$N_{\rm H}=3.87\times 10^{20} {\rm cm^{-2}}$ (Dickey \& Lockman, 1990).
The best fit result with a thermal plasma model is summarized in Table 1.
Since the source extent is small and is identical to point source for ASCA,  
all photons are fallen onto our selected region and we do not have to   
worry about the energy dependent ASCA PSF effect in the analysis (Takahashi et al. 1995).
The 2-10 keV band X-ray luminosity which is estimated from the above best fit values is
$4.9\times 10^{44} h^{-2}_{50} {\rm erg/sec}$.
However, a single power law model also gives an acceptable fit.
The best fit results are summarized in Table 2.
The best fit photon index of $1.8$ is typical of AGN's.
Therefore, one cannot distinguish if the source is a cluster or  AGN.

\subsection{ROSAT/HRI observation}
The ROSAT/HRI observation of CL2236-04 was performed in
December 1996 with an exposure time of 47ksec.
A diffuse source was detected at the position
of the cluster CL2236-04.
The background subtracted source count within a concentric circle centered
on the X-ray peak emission
increases monotonically up to $1'.4$ and it is
$481\pm 43$ source counts, within a circle of radius
$1^{\prime}.4$, which yields a source count rate of $(1.0\pm 0.1)\times 10^{-2}
{\rm cts/sec}$ where errors are $1\sigma$ Poisson error.
There is no other source with higher than $3\sigma$ detection significance
within a circle of radius $3'$ centered on the cD.
Therefore, we conclude that this source is identical to
AXJ2239-0429.
The HRI count rate is a factor of 1.4 higher than
 that expected from the ASCA observation,
that is $\sim 7\times 10^{-3}{\rm cts/sec}$.
Figure 2 shows the source image  obtained by ROSAT/HRI
superposed on the optical image of CL2236-04 obtained by ESO 3.6m (Melnick {\it et al.} 1993).
The maximum of the X-ray emission is consistent with the position
of the cD galaxy within the HRI pointing accuracy of a few arcsec.

Figure 3 shows the background subtracted
radial profile of AXJ2239-0429 obtained by ROSAT/HRI.
The radial profile is fitted to a, so called, {\it isothermal $\beta$ model} :
\begin{equation}
S_{{\rm X}}(\theta) = S_{0} \left(1 + \frac{\theta^2}{\theta_{c}^2}\right)^{-3\beta + \frac{1}{2}} 
\end{equation}
where $\theta_{c}$ is angular core radius and
best fit parameters are summarized in Table 3.
The source is clearly more extended than the PSF profile.
Therefore, we conclude that AXJ2239-0429 is
X-ray emission from the cluster of galaxies CL2236-04.
This is the first X-ray detection from the cluster.
Because redshift measurement has been done only for
a few member galaxies, the X-ray detection confirms
that CL2236-04 is a real gravitationally-bound entity.
In Table 4, the total gas mass and the total gravitational mass as calculated below are
summarized. The gas mass fraction is typical value.

The central electron number density gives a cooling time of 3.6 Gyr (from eq.(5.23) of Sarazin 1986),
which is about a half the age of the universe at $z = 0.552$.
This short cooling time and
the higher count rate of ROSAT/HRI compared to ASCA
may be a signature of cooling flow.
The cooling radius at where the cooling time equals to the age of the 
universe at $z=0.552$, is $\sim 10".6=78.4$kpc.
About $30\%$ of the total emission detected  by HRI is coming from the region within 
this cooling radius.
Since the  extent of the cooling flow is too small for the ASCA PSF  to resolve 
the cooling flow region
and the photon statistics of the ASCA spectra of AXJ2239-0429 is 
too poor to introduce further parameter, 
the current ASCA spectra can not constrain the cooling flow model. 
\section{Comparison with lens models}
\subsection{Spherically Symmetric Lens Model}
Firstly, we examine if the cluster can be modeled by a spherically
symmetric mass distribution.
Assuming that the X-ray emitting hot gas is isothermal and in
hydrostatic equilibrium, the hydrostatic equilibrium equation gives the cluster mass
contained within a radius $r$:
\begin{equation}
M(r)={kT\over G \mu m_{\rm H}} 3\beta {r^3\over r_c^2}{1\over
1+({r\over r_c})^2}
\end{equation}
where $\mu=1.3/2.1$ is the mean molecular weight for gas of cosmic
abundance.
Hereafter, this model is referred to as spherical isothermal $\beta$ model.
The lens equation for this mass distribution is
\begin{equation}
\tilde\theta_S= \tilde\theta_I - D {\tilde\theta_I \over
{\sqrt{\tilde\theta_I^2 + 1}}}
\end{equation}
where
\begin{equation}
D \equiv{{6 \pi \beta} \over {\theta_c}} {{kT} \over
{\mu m_H c^2}}{{D_{LS}}\over{D_{OS}}},
\end{equation}
$\tilde\theta_S$ ($
\tilde\theta_I$) is the angle between the lens and the source (image)
in units of $\theta_c$, and $D_{LS}$ ($D_{OS}$) is the angular
diameter distance between the lens (the observer) and the source
(the explicit expression is given in Fukugita {\it et al.} 1992).
$D$ is called the lens parameter and determines the efficiency of the
lens.
When $D\le 1$, the lens always produces a single image and
image deformation is small.
On the other
hand, when $D > 1$,
equation (2) has three routes for $\tilde\theta_S=0$ such as
\begin{equation}
\tilde\theta_I = 0,\;\;\pm \sqrt{D^2-1},
\end{equation}
and an infinitely amplified circular image is
formed with a radius of $\sqrt{D^2-1}\times \theta_c$ which is
called the Einstein ring radius.
The tangentially stretched giant
luminous arc
is formed near the tangential critical line in a spherical symmetric
lens model. Therefore, a separation between the giant luminous arc
and the center of the cluster must be close to the Einstein ring radius.
Assuming that the giant luminous arc in CL2236-04 is appeared 
at the Einstein ring radius, 
the temperature and core radius measured by
X-ray observations are compared in Fig.4 with
a relation of those parameters
required to explain the existence of the giant luminous arc
at the observed position.
If the universe does not contain a cosmological constant,
a mass distribution model
derived from X-ray observations with a
spherical isothermal $\beta$ model can not explain the existence of the giant luminous
arc at the observed position
even if 90\% one parameter errors in temperature
and core radius are taken into account.
The observation of the giant luminous arc requires a higher temperature
and smaller core radius, that is, it requires  a stronger mass concentration
in the cluster center
than that derived from the X-ray observations.
As shown in Figure 4,
an introduction of a cosmological constant implies that smaller mass concentration
is able to reproduce the same Einstein ring radius compared to a zero cosmological constant
model (e.g. Hamana {\it et al.} 1997).
The temperature and core radius relation derived from the giant luminous arc
position
with $(\Omega_0,\;\Lambda_0)=(0.1,0.9)$ has an overlapping region with
90\% error box obtained by X-ray observations.
\subsection{Non-spherical and clumpy Lens Model}
There are several pieces of evidence indicating that
deviations of the mass distribution in CL2236-04 from spherical symmetry
are playing significant roll in the
formation of the observed giant luminous arc.
First of all, the almost straight nature of the giant luminous arc
and  the fact that no counter image candidate has been found in spite of
a deep and wide field search (Kneib {\it et al.} 1994),
can not be explained by a fold arc produced by a spherical symmetric
lens.  It implies either an elliptical or substructural nature
of the lens (Narasimha \& Chitre  1993; Kneib {\it et al.} 1994).
Secondly, the X-ray morphology shows a clear
deviation from a circular symmetry and elongation in the
North-East to the South-West direction.
Finally, there is a giant  galaxy very close to the giant luminous
arc and a lensing effect by this galaxy should not be negligible.

Currently, the lens model proposed by
Kneib, Melnick \& Gopal-Krishna (1994; hereafter KMG) is
the only successful model which is consistent
with observed lensed image features.
The model assumes
two clumps of mass in this cluster.
It predicts that the core radius
($r_c$) and the 1D central velocity dispersion
($\sigma_0$) for the two clumps are $r_c=7''$(52$h_{50}^{-1}$kpc),
 $\sigma_0= 610{\rm km\;s^{-1}}$ for the first clump centered on the
cD galaxy,
and $r_c=3''$(22$h_{50}^{-1}$kpc), $\sigma_0= 390{\rm km\;s^{-1}}$ for
the second clump centered on the second giant galaxy in the cluster where
$H_0=50h_{50}{\rm km\;s^{-1}\;Mpc^{-1}}$ and $\Omega_0=1$.
The iso-mass densities of this model are shown in Fig.6 in KMG
(note south is up in the figure.).

If a  hydrostatic distribution for the X-ray emitting  gas
is assumed (for validity of this assumption see  e.g.
Schindler 1996), this lens mass-model
can be directly tested by X-ray observations.
First of all, the coincidence between the X-ray peak
and the cD center indicates that the center of
cluster gravitational potential coincides with the
cD position as predicted by KMG.
The  core radius predicted by the KMG is
consistent with the ROSAT/HRI result within
errors.
The position angle of the X-ray image 
surprisingly coincides with the prediction of KMG.
In the north of the cD galaxy, 
the direction of the major axis of the X-ray image is tilted toward
north. 
It might be evidence of substructure in this direction 
(Neumann \& B\"ohringer 1997) and
is roughly consistent with the position of the substructure 
predicted by the KMG.
In the limit of zero ellipticity, the lens equation of the cluster
adopted by KMG (Mellier, Fort \& Kneib 1993) 
has the same functional form as Eq.(3). 
Since $\beta (kT/\mu m_H)/\theta_c$ in Eq.(3) ($\sim 5.5\times 10^{14}{\rm  cm^2/s^2/arcsec}$) 
and the measured core radius are almost the same as $\sigma_0^2/r_c$ in KMG model 
($\sim 5.3\times 10^{14}{\rm  cm^2/s^2/arcsec}$) and KMG's core radius within the errors, 
the potential depth of the KMG model of cluster mass distribution is 
consistent with that inferred from the X-ray observations under assumptions of 
isothermal and hydrostatic equilibrium.
In Fig.3, it is shown the best fitting surface brightness model assuming that the
hot gas is isothermal with the best results for ASCA (dashed-line)
and the 90\% lower limit for ASCA (dashed-dotted line), 
and hydrostatic in the gravitational potential 
of the KMG model cluster with zero ellipticity and no substructure.   
In this fitting, only the central surface brightness is treated as free
fitting parameter.  Note that the 3D cluster mass distribution is deduced 
as shown in Mellier et al. (1993) in this procedure.
Although the model profile with the best fit temperature is too flat,  
the profile with the 90\% lower limit temperature gives a reasonably 
good fit where the fitting result yield $\chi^2/{\rm d.o.f.} = 73.65/59 = 1.25$. 
These results show that the KMG model is surprisingly consistent with the
the results of the X-ray observations. 
\section{Discussions and Conclusion}
A  new X-ray source in the direction of
the distant lensing cluster, CL2236-04,
is discovered by ASCA, named AXJ2239-0429.
The source extent is resolved by the following ROSAT/HRI observation
and it confirms that the X-ray source is
the cluster of galaxies, CL2236-04.

The mass distribution of the cluster derived by  X-ray observations
and by the observed nature of the giant luminous arc is
compared.
It is found that within the frame work of  a spherical isothermal $\beta$ model,
the required mass  to explain the observed location of the
giant luminous arc is larger than that expected from the 
X-ray observations as reported for 
many giant luminous arc
clusters (Wu \& Fang 1997).
An introduction of  a large cosmological constant reduces
the lensing mass but the total reduction  is only 20-30\%.
Therefore, the cosmological constant can not be the 
main solution for the discrepancy found in many clusters. 

Cooling flow can be one of the other possible solutions (Allen, Fabian \& Kneib 1996, 
Allen 1997). 
According to the cooling flow model, the hot gas in the
cluster central region is in the multi-temperature phase.
The temperature of the hottest phase gas which represents a real 
gravitational potential depth, obtained by the multi-phase model 
fitting becomes much higher than the temperature 
obtained by the single phase model fitting because the last one
is the average temperature of the multi-temperature gas. 
Allen (1997) is claiming that the most of
the discrepancy reported before can be explained by the 
existence of the cooling flow.   
There are several evidences of the existence of 
the cooling flow in the central region of  AXJ2239-0429 as 
shown in Sec.2. 
It may be possible that the measured temperature of AXJ2239-0429 underestimates
the cluster potential depth.
If the ASCA temperature underestimates the cluster potential depth by a factor 
of two due to the cooling flow, 
the Einstein ring radius of the cluster for a source at $z=1.116$ becomes as large as 
the distance  between the giant luminous arc and the cluster center. 
However, the straight nature of the arc and a lack of the bright 
counter image can not be explained by a spherically 
symmetric lens.  
Further, the perturbation by the gravitational potential of galaxy A is 
not negligible. 
The absolute luminosity of galaxy A in visual band is $3.3\times 10^{11}h_{50}^{-2}L_{\odot}$
(Melnick et al. 1993).   This luminosity is translated into the 
the line of sight velocity dispersion of $\sim 370{\rm km/s}$ by using 
the Faber-Jackson law (Binney \& Tremaine 1987).   
If the potential depth is about a factor of two deeper than that inferred from the ASCA 
temperature, it is impossible to construct the model which is able to 
explain the location and the observed feature of the giant luminous arc due to 
the large perturbation of the galaxy A.
Therefore, we conclude that even if the AXJ2239-0429 has a cooling flow in the 
central region, the temperature measured by ASCA which is the mean emission-weighted cluster
temperature in this case, should not be cooler than and different from the virial temperature
of the cluster. 

The most plausible solution is effect of the clumpiness of the 
gravitational potential due to galaxy A 
and non-zero ellipticity in the mass distribution of the cluster.
KMG  proposed the cluster mass distribution model 
by taking into account these effects which explain the 
observed features of the giant luminous arc. 
The ellipticity, position angle and velocity dispersion
of the  cluster predicted by KMG is surprisingly
consistent with the results of the X-ray observations of the cluster
as shown in the previous section. 
The velocity dispersion of galaxy A assumed by KMG is also  
consistent with that deduced by the Faber-Jackson law.
Therefore, we conclude that the effect of the clumpiness of the 
gravitational potential due to galaxy A 
and non-zero ellipticity in the mass distribution of the cluster 
are essential to explain the observed feature of the giant luminous
arc and the cluster mass distribution required by the
strong lensing effect is totally consistent with 
the X-ray observations assuming that the X-ray emitting hot gas is
isothermal and in a hydrostatic equilibrium state.   
\acknowledgements
The authors would like to thank J. Melnick, B. Altieri, Gopal-Krishna,
E.Giraud for providing the optical image of CL2236-04.
MH and KM would like to thank Takashi Hamana and Mikio Murata
for valuable discussions.
M.H. is supported by Yamada Science Foundation and by the
Grants-in-Aid by the Ministry of Education, Science and Culture of 
Japan (09740169, 60010500).
%
%
\newpage
%
 
%
%
\newpage
\figcaption{X-ray spectra of AXJ2239-0429 obtained by ASCA.
The ASCA observatory has four identical telescopes
covering the energy range of 0.3 to 10keV (Tanaka {\it et al.} 1994).
They are equipped with two gas imaging spectrometers (GIS 2 and GIS 3) and
two solid imaging spectrometers (SIS 0 and 1) at the focal plane.
The solid lines are the best fit Raymond-Smith models obtained by simultaneous
fitting of the GIS and SIS spectra.
\label{fig1}}
\figcaption{Significance contour plot of X-ray image of AXJ2239-0429
obtained by ROSAT/HRI (Tr\"umper 1984).
Subtracted background level is $1.25_{-0.01}^{+0.05}\times10^{-6} {\rm count\cdot sec\cdot arcsec^{-2}}$.
The step size is $2\sigma$. The lowest contour level is $2\sigma$.
This image is smoothed with a Gauss filter with a $\sigma$ of $15$ arcsec.
The superposed is V+R image of CL2236-04 taken by ESO 3.6m 
(Melnick et al. 1993).
\label{fig2}}
\figcaption{Background subtracted radial surface brightness profile of AXJ2239-0429 
obtained from the ROSAT/HRI data. 
Errors are $1\sigma$ and the solid line is the best fit with the parameters shown in Table 3. 
A  dotted line is a PSF profile at the X-ray peak. 
The best fitting surface brightness model assuming that the
hot gas is isothermal with the best fit temperature (dashed-line)
and the 90\% lower limit temperature for ASCA (dashed-dotted liine), 
and hydrostatic in the gravitational potential 
of the KMG model cluster, is also shown.  \label{fig3}}  
\figcaption{Comparison between the mass distribution
derived from  X-ray observations and
that from a giant luminous arc.
Bold lines show loci where the Einstein ring radii for the sources at $z=1.116$ equal
to a separation between the giant luminous arc in CL2236-04
and the center of cD galaxy, that is $12^{\prime\prime}.5$, for various cosmological
models, the solid line for $(\Omega, \Lambda)=(1.0,0)$,
the dashed line for $(\Omega, \Lambda)=(0.1,0)$, and the dot-dash-line for
$(\Omega, \Lambda)=(0.1,0.9)$ where the best fit value is
used for $\beta$ and a isothermal spherical symmetric
lens model is assumed.
In each cosmological model, the Einstein ring radius
is larger than $12^{\prime\prime}.5$ in the region above the line.
A region enclosed by 90\% measurement error
for temperature and core radius is shown by a dotted box.
The best fit values of temperature and core radius is shown by
a diamond.
As  a reference, loci where the lens becomes critical, $D=1$,
are shown by thin lines for each cosmological models.
At least the temperature and the core radius of the cluster must
be in the region  above these lines for the cluster to be
able to make strong lensing event for the sources at $z=1.116$.\label{fig4}}
%
\newpage
\begin{deluxetable}{ccccc}
\tablewidth{0pc}
\tablecaption{ASCA spectral fitting results with a thermal plasma model. Errors are 90\% confidence for one interesting parameter.}
\tablehead{
\colhead{$k_{\rm B}T_{{\rm X}}$} & \colhead{$Z_{\rm Fe}$} & \colhead{$N_{\rm H}$} & \colhead{$z$} & \colhead{$\chi^2$/d.o.f.}\\
\colhead{(keV)} & \colhead{($Z_{\odot}$)} & \colhead{($10^{20}{\rm cm^{-2}}$)} & \colhead{} & \colhead{}}
\startdata
$6.2^{+2.6}_{-1.7}$ & $0.0_{-0.0}^{+0.38}$ & $3.87$\tablenotemark{*} & $0.552$\tablenotemark{*} & $72.66/80=0.94$ 
\enddata
\tablenotetext{*}{fixed}
\end{deluxetable}
\begin{deluxetable}{ccc}
\tablewidth{0pc}
\tablecaption{ASCA spectral fitting results with a power low model}
\tablehead{
\colhead{Photon Index} & \colhead{$N_{\rm H}$} & \colhead{$\chi^2$/d.o.f.}\\
\colhead{} & \colhead{($10^{20}{\rm cm^{-2}}$)} & \colhead{}}
\startdata
$1.81_{-0.15}^{+0.15}$ & $3.87$\tablenotemark{*} & $75.00/80=0.96$
\enddata
\tablenotetext{*}{fixed}
\end{deluxetable}
\begin{deluxetable}{cccc}
\tablewidth{0pc}
\tablecaption{X-ray properties of AXJ2239-0429 from the spatial analysis of ROSAT/HRI data}
\tablehead{
\colhead{$S_{0}$} & \colhead{$\theta_{c}$} & \colhead{$\beta$} & \colhead{$\chi^2$/d.o.f.}\\
\colhead{(count/s/${\rm arcsec}^{2}$)} & \colhead{(arcsec)} & \colhead{} & \colhead{}}
\startdata
$1.10^{+0.30}_{-0.25}\times 10^{-5}$ & $11.39^{+5.5}_{-3.4}$ 84.2$h_{50}^{-1}$kpc& $0.66^{+0.22}_{-0.11}$ & $67.665/56=1.21$ 
\enddata
\end{deluxetable}
\begin{deluxetable}{lcc}
\tablewidth{0pc}
\tablecaption{The gas mass and gravitational mass of AXJ2239-0429 within $1'.4$($0.6h_{50}^{-1}$Mpc) and $1h_{50}^{-1}$Mpc}
\tablehead{
\colhead{} & \colhead{$0.6h_{50}^{-1}$Mpc} & \colhead{$1.0h_{50}^{-1}$Mpc}}
\startdata
$n_{e_{0}}$ ($h_{50}^{1/2}{\rm cm^{-3}}$) & \multicolumn{2}{c}{$2.44\times10^{-2}$}\nl
$M_{\rm gas}$ ($h_{50}^{-5/2}M_{\odot}$) & $3.26\times10^{13}$ & $5.76\times10^{13}$\nl
$M_{\rm grav}$ ($h_{50}^{-1} M_{\odot}$) & $2.69\times10^{14}$ & $4.39\times10^{14}$\nl
$M_{\rm gas}/M_{\rm grav}$ ($h_{50}^{-3/2}$ )& 0.12 & 0.13
\enddata
\end{deluxetable}
%
%
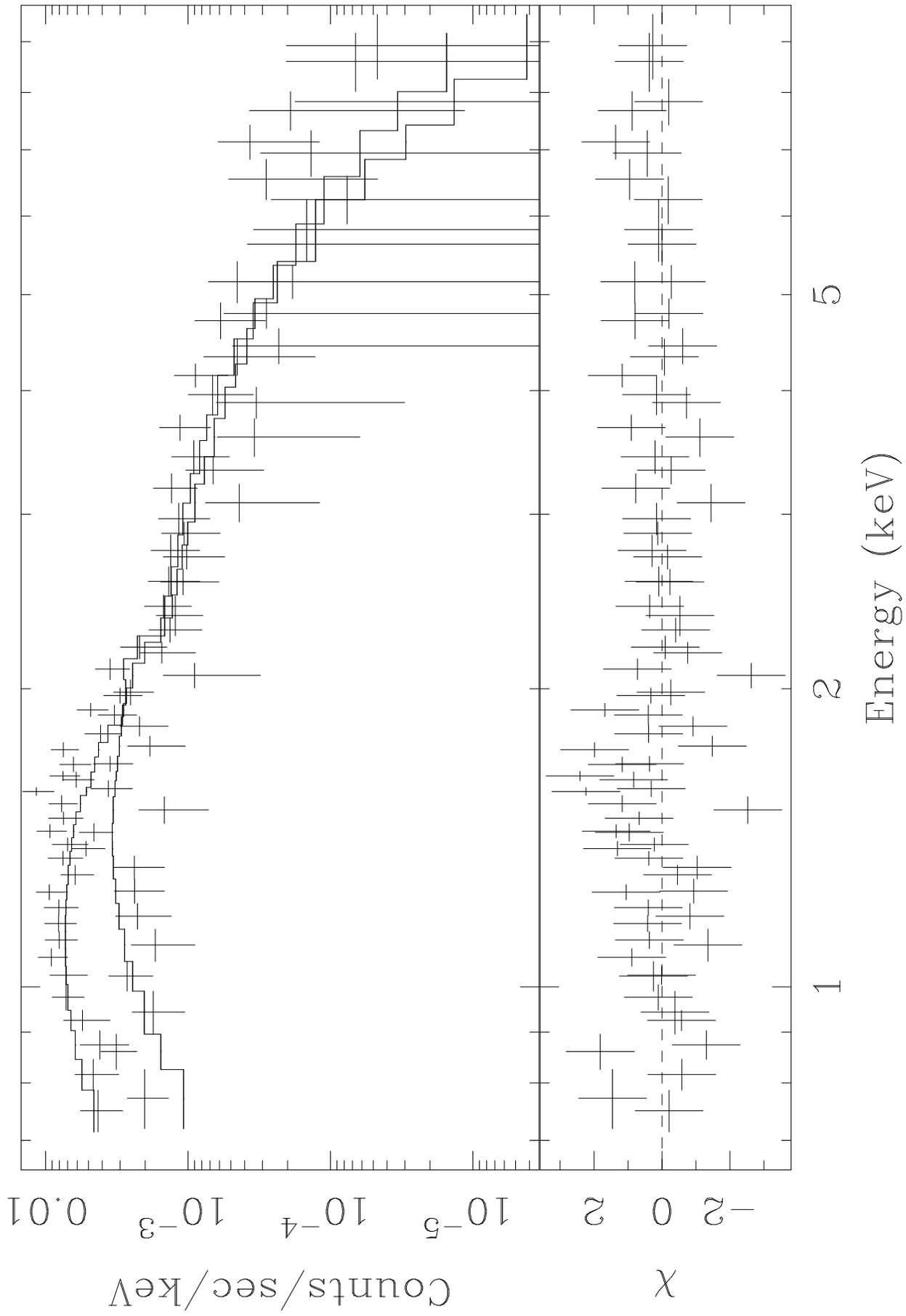
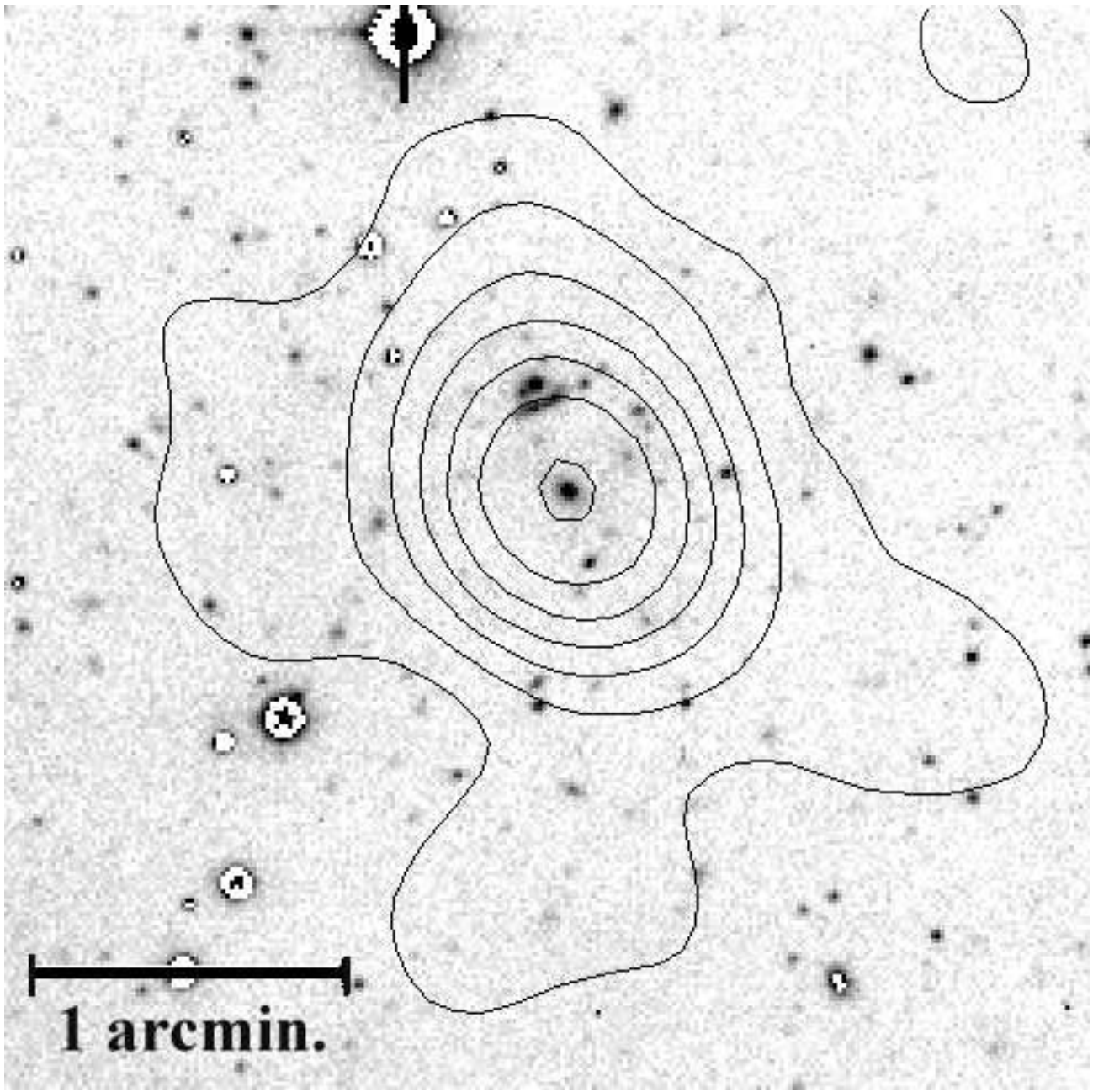
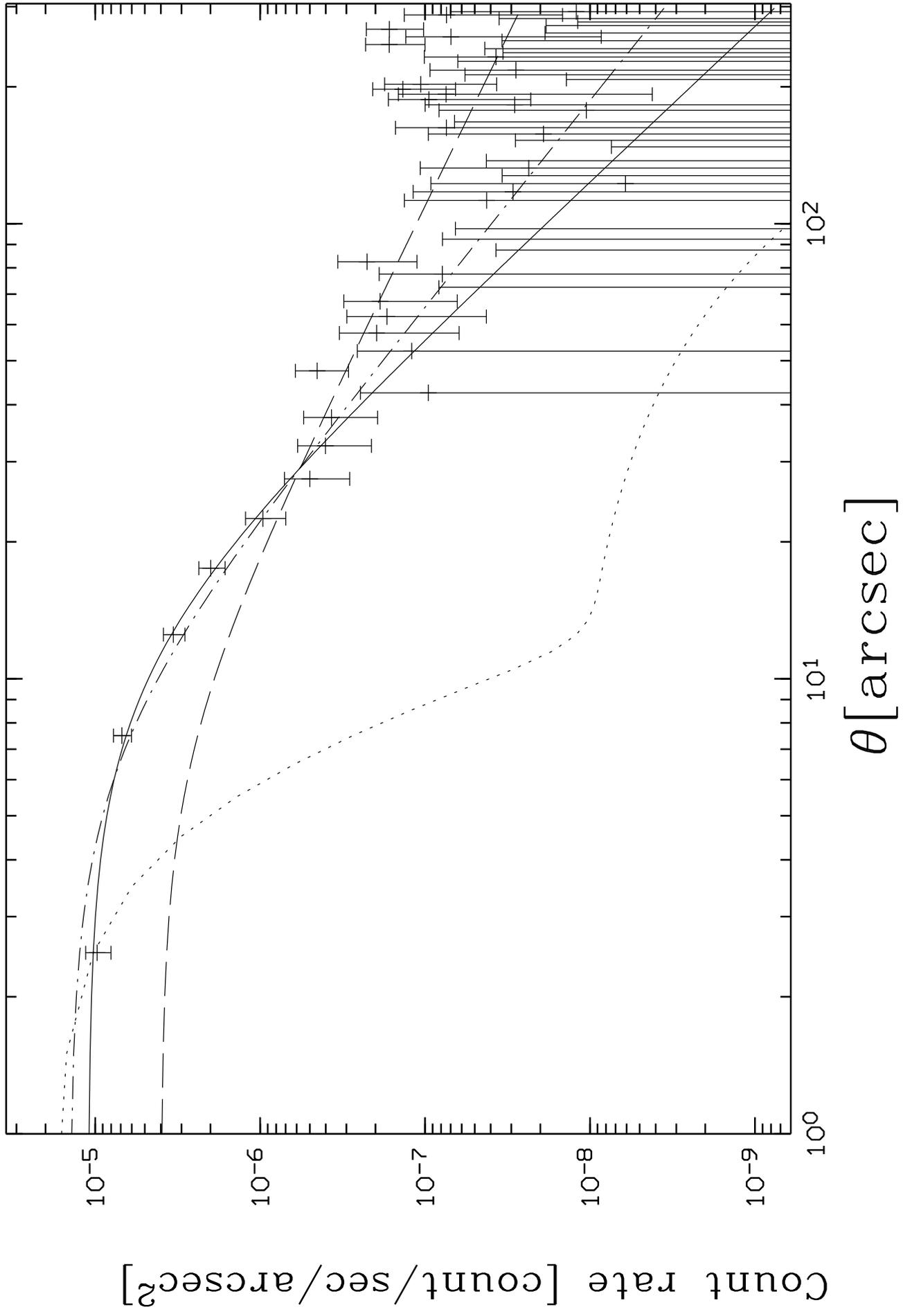
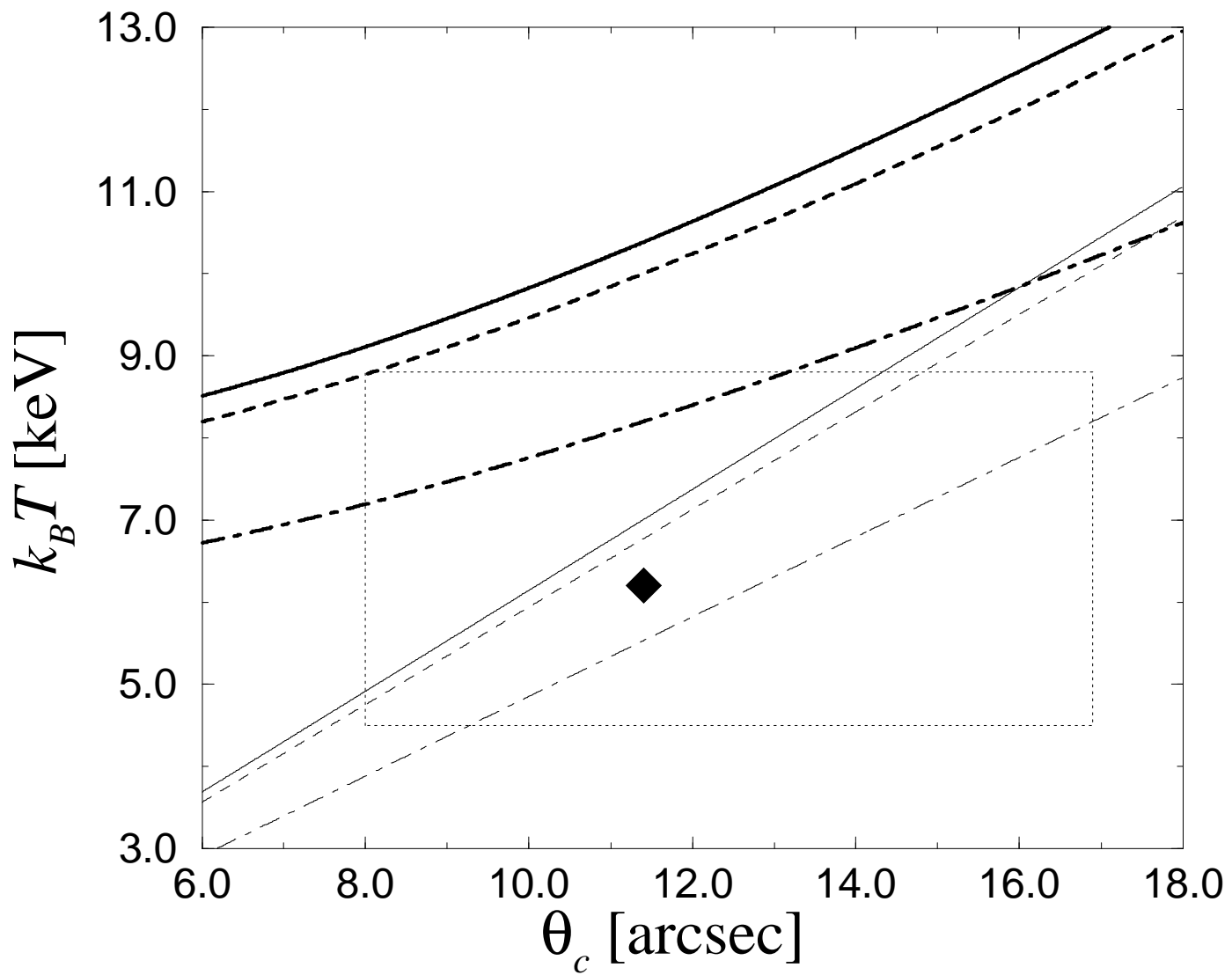
\end{document}